\documentclass[aps,prresearch,twocolumn,superscriptaddress,showpacs,amsmath,amssymb,floatfix]{revtex4-2}
\usepackage{lmodern}
\usepackage{graphicx}
\usepackage{bm}
\usepackage{dcolumn}
\usepackage[colorlinks=true,linkcolor=blue,citecolor=blue,urlcolor=blue]{hyperref}
\graphicspath{{figures/}}

\begin{document}

\title{Quantum jump flux guides adaptive-rank variational simulation of open systems}

\author{Yu-Hao Chen}
\affiliation{Department of Physics, National Taiwan University, Taipei, Taiwan}
\affiliation{Department of Artificial Intelligence, Chang Gung University, Taoyuan, Taiwan}
\author{Renata Wong}
\thanks{Corresponding author: Renata Wong (renata.wong@cgu.edu.tw)}
\affiliation{Department of Artificial Intelligence, Chang Gung University, Taoyuan, Taiwan}
\affiliation{Department of Neurology, Chang Gung Memorial Hospital, Keelung, Taiwan}
\affiliation{Artificial Intelligence Research Center, Chang Gung University, Taoyuan, Taiwan}

\date{\today}

\begin{abstract}
Low-rank variational simulation of Lindblad dynamics represents the
state as a mixture of basis states rotated by a shared quantum
circuit. The number of states needed can change during evolution.
We use quantum-jump flux to decide when to expand the mixture and
which state to add. At fixed circuit parameters, adding a state
with zero initial weight reduces the minimum squared McLachlan
residual by exactly the square of the jump flux into that state.
The square of the total flux into omitted states bounds the
improvement from adding any subset of them. 
When leakage is large enough relative to both the trace and the
residual, we add the state with the largest incoming jump flux.
We remove states with low weight and little incoming flux, and
cap the total weight removed. Removing one state and renormalising
changes the mixture by a trace distance equal to its weight fraction.
Our trace-distance bound accounts for both the accumulated residual
and changes to the mixture. In 4-10-qubit pulse simulations, fixed
rank with a hindsight basis used 1.9-2.1 times as many measurement
settings at matched time-averaged trace distance.

\end{abstract}

\maketitle
\lefthyphenmin=3\relax
\righthyphenmin=3\relax

\section{Introduction}

The Lindblad master equation describes the dynamics of a wide class
of open quantum systems~\cite{lindblad1976,gks1976,breuer2007}.
The density matrix of an $n$-qubit system has $4^n$ entries, so
direct simulation quickly becomes expensive~\cite{weimer2021}.
One way to avoid evolving the full matrix is to simulate an
ensemble of quantum trajectories, each represented by a state
vector~\cite{dalibard1992,plenio1998,daley2014}. Other classical
approaches use tensor networks~\cite{verstraete2004,zwolak2004},
corner-space renormalisation~\cite{finazzi2015}, or neural-network
ans\"atze~\cite{nagy2019,vicentini2019,hartmann2019,yoshioka2019}.

Quantum approaches include fault-tolerant
algorithms~\cite{kliesch2011,childs2017,cleve2017} and near-term
schemes based on dilation or operator
decomposition~\cite{sweke2015,hu2020,schlimgen2021,delgado2025}.
Variational quantum simulation uses McLachlan's principle to choose
parameter velocities that minimise the mismatch with the target
dynamics~\cite{mclachlan1964,li2017,yuan2019,mcardle2019,cerezo2021,bharti2022}.
Several open-system formulations encode the vectorised density
matrix on $2n$ qubits~\cite{endo2020,yoshioka2020,watad2024,chen2024,kamakari2022}.
In the representation of Santos, Song and Savona~\cite{santos2025},
one parameterised circuit on $n$ qubits rotates $R$ orthogonal
computational-basis states. A classical list stores their mixture
weights. The \emph{represented support} is the set of $R$ labels
whose weights the simulation tracks, including any that are
currently zero. If $R$ is too small, dissipation can populate
states outside that support. Keeping more states costs more
measurements, and very
small weights can make the variational equations poorly conditioned.
The right support size can therefore change as the system evolves.

Santos, Song and Savona~\cite{santos2025} linked loss of trace to jumps
out of the represented support and proposed adapting it during
evolution. We call the rate at which jumps feed weight into a given
state its incoming \emph{jump flux}. The sum of those rates over
all omitted states is the \emph{trace leakage}. Classical low-rank
methods use error estimates to adapt
the represented subspace~\cite{lebris2013,lebris2015,gravina2024}.
Related variational approaches add circuit
generators~\cite{chen2024} or select a rank by spectral
truncation~\cite{zeynali2026}.

At each scheduled check, a two-part rule called the
\emph{balanced trigger} decides whether to look for states to
add. The first test compares the leakage with the weight
still represented. The second compares it with the residual,
the mismatch between the variational and Lindblad time
derivatives. If the leakage is small relative to that
mismatch, adding states can do little to improve the fit.
If both tests pass, we consider candidates in order of their
incoming jump flux. We also allow low-weight states to leave
the support, with a limit on how much weight can be
discarded over a run.

Classical Lindblad error bounds account for truncation and
time-integration errors~\cite{etienney2026}, and
Ref.~\cite{jassem2026} studies residuals and discarded weight in
spectral compression. Our \emph{trace-distance certificate}
is an upper bound on the trace distance between the normalised
variational state and the exact Lindblad state. It uses the
residual accumulated during evolution and the size of each
discrete update. The same mixture weights that guide changes to
the support also help us allocate measurement shots when
estimating the variational equations.

\section{Theory and adaptive protocol}\label{sec:methods}

\subsection{Closed-form rank identities}\label{sec:ansatz}

Let $S=S(t)$ be the set of retained computational-basis labels at
time $t$ and $R=|S|$. The variational state is
\begin{align}
\rho(\alpha,\theta)&=\sum_{x\in S}\alpha_xP_x,\nonumber\\
P_x&=|\psi_x\rangle\langle\psi_x|,\qquad
|\psi_x\rangle=U(\theta)|x\rangle .
\end{align}
Here $U(\theta)$ is a shared circuit of Pauli rotations with gate
angles $\theta$, and $\alpha=(\alpha_x)_{x\in S}$ contains the
nonnegative mixture weights. Its trace is
$\tau=\operatorname{Tr}\rho=\sum_{x\in S}\alpha_x$.
A \emph{zero-weight component} has a retained label $x\in S$
and a tracked weight $\alpha_x=0$. Its term $\alpha_xP_x$
contributes zero to the current density matrix, and its weight
can grow in later updates. An \emph{omitted} label $x\notin S$
gets a weight coordinate only when it is added to $S$.
The support size $R$ counts every retained label. The matrix
rank $R_a$ counts only the components with positive weight.

The target evolution is generated by the Lindblad operator
\begin{equation}\label{Lindblad}
\mathcal L[\rho]=-i[H,\rho]+\sum_{k=1}^K\gamma_k
\Big(c_k\rho c_k^\dagger-\tfrac12\{c_k^\dagger c_k,\rho\}\Big),
\end{equation}
where $H$ is the Hamiltonian, $K$ is the number of jump operators,
and $c_k$ and $\gamma_k$ are the $k$th jump operator and its rate.
We set $\hbar=1$. Square and curly brackets denote the commutator
and anticommutator.

Let $\beta=(\alpha,\theta)$ and $\partial_i\rho=\partial\rho/\partial\beta_i$.
McLachlan's principle chooses velocities $\dot\beta$ to minimise
$\|\dot\rho-\mathcal L[\rho]\|_2$, with
$\dot\rho=\sum_i\dot\beta_i\partial_i\rho$.
The Hilbert--Schmidt norm is
$\|A\|_2=\sqrt{\operatorname{Tr}(A^\dagger A)}$.
The normal equations are $M\dot\beta=V$, where
\begin{align*}
M_{ij}&=\operatorname{Tr}(\partial_i\rho\,\partial_j\rho),\nonumber\\
V_i&=\operatorname{Tr}(\partial_i\rho\,\mathcal L[\rho]).
\end{align*}
Orthonormality gives $M_{\alpha\alpha}=I$ and $M_{\alpha\theta}=0$,
where $I$ is the identity on the weight coordinates. Thus
\begin{align}
\dot\alpha_x&=\langle\psi_x|\mathcal L[\rho]|\psi_x\rangle,
\label{eq:weightvelocity}\\
M_{\theta\theta}\dot\theta&=V_\theta,\nonumber\\
D^2&=\|\mathcal L[\rho]\|_2^2-\sum_{x\in S}\dot\alpha_x^2
      -V_\theta^T M_{\theta\theta}^{+}V_\theta.\nonumber
\end{align}
The residual is the mismatch between the variational time
derivative and the Lindblad time derivative at the same state.
Here $D=D_S$ is the smallest norm of that mismatch on $S$.
The matrix $M_{\theta\theta}^{+}$ is the Moore--Penrose pseudoinverse of the
angle block, and $V_\theta$ is the angle part of $V$. The rank
identities use this unregularised optimum. The certificate uses
the residual $r=\dot\rho-\mathcal L[\rho]$ from the velocities
actually applied, with $D_{\rm app}=\|r\|_2$. A regularised
solve can give $D_{\rm app}>D$.

\emph{State addition.} Adding an omitted label $c\notin S$ with
zero initial weight leaves $\rho$ unchanged. The new weight gives
the time derivative one more independent direction. It can now
start putting weight on $|\psi_c\rangle$. This direction is
orthogonal to the existing variational directions. Since
$\rho|\psi_c\rangle=0$, only jumps contribute along that direction:
\begin{align}
V_c&=\sum_{k=1}^K\gamma_k\sum_{a\in S}
       \alpha_a|\langle\psi_c|c_k|\psi_a\rangle|^2\ge 0,\nonumber\\
D^2_{S\cup\{c\}}&=D^2_S-V_c^2 .
\end{align}
At the current state and circuit angles, the largest incoming
flux $V_c$ gives the largest drop in squared residual. We can
therefore rank candidates before adding any of them.

\emph{Trace leakage.} When the weights evolve according to
Eq.~\eqref{eq:weightvelocity}, completeness of the rotated basis
gives the trace-leakage identity~\cite{santos2025}
\begin{equation}
\Gamma_{\rm leak}=-\dot\tau=\sum_{c\notin S}V_c .
\end{equation}
The projected weight equations track only the retained
components, so flux into omitted states appears as a loss of
trace in $\rho$. The exact Lindblad evolution preserves the trace.

\emph{Residual splitting.} The zero-weight directions are
mutually orthogonal. Let $D_{\rm circuit}$ be the minimum residual
norm when all basis labels are available at the same state and
gate angles. Then
\begin{align}
D^2&=D_{\rm circuit}^2+\sum_{c\notin S}V_c^2,\nonumber\\
\sum_{c\notin S}V_c^2&\leq\Gamma_{\rm leak}^2.
\end{align}
We call the sum of squared fluxes the rank contribution because
it is the part of $D^2$ that adding basis states can remove at
the current angles. The remaining $D_{\rm circuit}^2$ is the
circuit contribution.
In particular,
\begin{equation}
D_{\rm circuit}^2\geq D^2-\Gamma_{\rm leak}^2.
\label{eq:rankbound}
\end{equation}
Thus $\Gamma_{\rm leak}^2$ bounds the improvement in $D^2$ from
any support expansion at fixed angles. For a positive threshold
factor $\kappa$, if $\Gamma_{\rm leak}<\kappa D$, then
\begin{equation}\label{eq:kappabound}
D_{\rm circuit}^2>(1-\kappa^2)D^2 .
\end{equation}
Even adding the entire missing support would then reduce $D^2$
by less than the fraction $\kappa^2$. The bound can be loose
when the leakage is spread over many candidates.

\emph{State removal.} Write $\lambda_x=\alpha_x/\tau$ for the
normalised weight of component $x$. If $\lambda_x<1$, removing
$x$ and renormalising changes the normalised state by trace
distance $\lambda_x$. The two states share the rotated basis,
so this is the single-component discarded-weight
identity~\cite{jassem2026}. For states $\sigma$ and $\rho$,
the trace distance is
$\mathrm{TD}(\sigma,\rho)=\tfrac12\|\sigma-\rho\|_1$, where
$\|A\|_1$ is the sum of the singular values of $A$.
Appendix~\ref{app:a} derives the four rank identities.

\emph{Rate equation.} The Hamiltonian contributes zero to the
diagonal weight derivative. The jump terms give
\begin{equation}\label{eq:rate}
\dot\alpha_x
=\sum_{a\in S}W_{a\to x}\,\alpha_a
-\alpha_x\sum_y W_{x\to y}.
\end{equation}
The transition rate from $i$ to $j$ is
$W_{i\to j}=\sum_{k=1}^K\gamma_k
|\langle\psi_j|c_k|\psi_i\rangle|^2$~\cite{dalibard1992,plenio1998}.
The first sum is inflow from retained states. The second is
outflow to the full rotated basis, including states outside $S$.

To estimate the rates in Eq.~\eqref{eq:rate}, we can sample the
destination of a jump from a retained state. Prepare
$|\psi_a\rangle$ and implement the jump operator $c_k$ using an
ancilla or, where available, a dynamic circuit~\cite{corcoles2021}.
On a successful jump, apply $U^\dagger$ and measure in the
computational basis. The probability of outcome $x$, conditional
on the jump, is
\[
p(x\mid a,k)=
\frac{|\langle\psi_x|c_k|\psi_a\rangle|^2}
     {\|c_k|\psi_a\rangle\|^2}.
\]
Multiplying by the jump rate $\gamma_k\|c_k|\psi_a\rangle\|^2$
and summing over $k$ gives $W_{a\to x}$. We use these rates with
the stored mixture weights to estimate leakage, rank candidates
for addition, and check the incoming flux before removing a state.

\subsection{Adaptive protocol}\label{sec:protocol}

We start with the reference bitstring ($R=1$) and zero gate
angles ($\theta=0$). The fourth-order Runge--Kutta method (RK4)
advances the weights and angles. The gate order stays fixed
throughout. A \emph{checkpoint} is a step at which we check
whether to add or remove labels.

\begin{enumerate}
\item \textbf{Schedule a checkpoint.}
After each step, compute the von Neumann entropy
$S_{\rm vN}=-\sum_{x\in S}\lambda_x\ln\lambda_x$ of the normalised
state, with $0\ln0=0$. The scaled entropy is
$\eta=S_{\rm vN}/\ln R$ for $R>1$ and $\eta=0$ for $R=1$.
The weight equations also give
$\Gamma_{\rm leak}=-\sum_{x\in S}\dot\alpha_x$.
The default interval between checkpoints is 10 steps. We shorten
it to 2 steps if the entropy has increased since the previous step
or if $\eta>0.9$.

\item \textbf{Check whether to remove a state.}
A component is eligible if $\lambda_x<10^{-3}$ and
$\dot\alpha_x\leq0$. We add its normalised weight just before
deletion to a running total. A removal can go ahead only if
that total stays at or below $10^{-2}$ over the run.

We also check how much flux still enters the candidate from
the states we plan to keep. This check, called the \emph{inflow
guard}, keeps us from removing a state that would cause too much
additional leakage. It adds the incoming flux to
the leakage estimate, including the flux from any removals
already counted at that checkpoint. After dividing by the trace
before removal, the estimate must be at or below half the
addition threshold.

\item \textbf{Check whether to add states.}
The balanced trigger has two tests. First, the fractional
rate of trace loss, $\Gamma_{\rm leak}/\tau$, must exceed
$3\times10^{-3}$ per unit time. Dividing by $\tau$ measures
the leakage relative to the weight still represented.
Second, $\Gamma_{\rm leak}$ must be at least $\kappa D$,
where $D$ is the minimum residual norm and $\kappa=0.3$
by default. Equation~\eqref{eq:kappabound} gives the reason
for the second test. If it fails at the default $\kappa$,
adding even the entire missing support at
the current angles can reduce $D^2$ by less than 9\%.
Passing both tests starts the candidate search. Each
candidate must still meet the flux threshold below.

With finite shots, we estimate $D^2$ twice using independent
samples. We take half the absolute difference as a rough measure
of noise. If the mean exceeds three times that measure, we use
its square root for $D$ in the second test. Otherwise, we apply
only the leakage threshold.

\item \textbf{Find and add candidates.}
The Hamming distance between two bitstrings is the number of bit
positions at which they differ. Start with omitted labels at
distance one from any retained label. If candidates account for
less than 90\% of the total leakage, increase the maximum distance
one bit at a time, up to three bits by default. The jump circuit
in Sec.~\ref{sec:ansatz} can also sample candidates directly.

We consider candidates in decreasing order of incoming jump
flux $V_c$. We add up to two per checkpoint, each with
$V_c\geq 0.05\Gamma_{\rm leak}$, and initialise their weights to
zero. A previously removed state must wait at least 20 steps
and meet twice the usual candidate flux threshold.
\end{enumerate}

\subsection{Continuous-time trace-distance certificate}

When deciding whether to add a state, we need the residual at
the current checkpoint. To bound the error at time $t$, we need
the residual accumulated along the trajectory and the changes
made at every update. The theorem below shows how these two
contributions enter the certificate.

\emph{Theorem (continuous-time trace-distance certificate).}
Let $\rho^*(t)$ be the exact state under a Lindblad generator
$\mathcal L_t$ whose propagator is completely positive and
trace preserving. Let $\rho(t)$ be a positive semidefinite
variational state that is differentiable between discrete
updates, with $\rho(0)=\rho^*(0)$ and
$\tau(t)=\operatorname{Tr}\rho(t)>0$. Define
\[
r(t)=\dot\rho(t)-\mathcal L_t[\rho(t)]
\]
between updates, using the parameter velocities actually
applied. At an update time $t_j$, define
$\Delta_j=\rho(t_j^+)-\rho(t_j^-)$ and take $\rho(t_j)$ to
be the state just after the update. Then
\begin{equation}
\mathrm{TD}\!\left(\frac{\rho(t)}{\tau(t)},\rho^*(t)\right)
\leq \frac{1}{\tau(t)}
\left[
\int_0^t\|r(s)\|_1\,ds
+\sum_{t_j\leq t}\|\Delta_j\|_1
\right],
\label{eq:certificate}
\end{equation}
where $\|A\|_1$ is the trace norm and
$\mathrm{TD}(\sigma,\rho)=\tfrac12\|\sigma-\rho\|_1$.

Adding a component at zero weight leaves the state unchanged, so
that update contributes zero. Deleting component $x$ subtracts
$\alpha_xP_x$ from the unnormalised density matrix and contributes
$\alpha_x$ to the sum. If we clip negative weights at the end of
an accepted step, the contribution is
$\sum_x|\alpha_x^+-\alpha_x^-|$, with $+$ and $-$ denoting the
weights after and before clipping. Clipping at intermediate RK4
stages affects the time-stepping error.

Appendix~\ref{app:b} proves the theorem. It relies on
trace-norm contractivity for Hermitian operators
\cite{nielsen2010,ruskai1994}. An analogous argument in the
Hilbert--Schmidt norm~\cite{santos2025,zoufal2023} needs
contractivity in that norm. It holds for differences of states
under unital and single-qubit channels~\cite{perezgarcia2006}
and can fail for higher-dimensional non-unital
channels~\cite{wang2009}. Appendix~\ref{app:c} gives a
four-qubit example.

\emph{Evaluating the residual.}
In classical simulations we factor the residual as
$r=YGY^\dagger$, where $Y$ is a matrix of vectors and $G$
collects the coefficients of their outer products.
The columns of $Y$ are the $R$ represented state vectors and,
for each of the $R_a$ nonzero-weight components, the state
velocity $|\dot\psi_a\rangle$, the vector
$H_{\rm eff}|\psi_a\rangle$, and the $K$ vectors
$c_k|\psi_a\rangle$. The effective Hamiltonian between jumps is
$H_{\rm eff}=H-\tfrac{i}{2}\sum_k\gamma_k c_k^\dagger c_k$.
The column count gives the structural bound
$\operatorname{rank}(r)\leq\nu=\min\{2^n,R+(K+2)R_a\}$.

\emph{Bounding the residual from measurements.}
The singular values of $r$ give
\begin{equation}
\|r\|_1
\leq \sqrt{\operatorname{rank}(r)}\,\|r\|_2
\leq \sqrt{\nu}\,D_{\rm app}.
\label{eq:rankresbound}
\end{equation}
At a checkpoint, Appendix~\ref{app:b} expresses $r$ as a sum
of unit-vector outer products with coefficients $b_j$. It uses
two independent Hadamard tests per sample to construct a signed
outcome $Z\in\{-1,1\}$ with
$D_{\rm app}^2=2\Lambda_r^2\mathbb E[Z]$,
where $\Lambda_r=\sum_j|b_j|$ and $\mathbb E[Z]$ is the mean
over index draws and circuit outcomes. For $N$ independent
pairs, write $\overline Z$ for the sample mean and
$\widehat d=2\Lambda_r^2\overline Z$. For failure probability
$0<\delta<1$, Hoeffding's inequality gives, with probability
at least $1-\delta$,
\begin{equation}
D_{\rm app}^2
\leq \widehat d+
2\Lambda_r^2\sqrt{\frac{2\ln(2/\delta)}{N}}.
\label{eq:ducb}
\end{equation}

Equation~\eqref{eq:ducb} applies at one checkpoint. A trajectory
certificate also requires bounds on the residual integral
between checkpoints and on time-stepping error
(Appendix~\ref{app:b}).

\section{Benchmark setup}\label{sec:benchmarks}

We compare two ways of applying the transverse field. In the
\emph{pulse}, the field is applied for a finite interval and then
turned off, so we can see how the support grows during the drive
and shrinks as the system relaxes. In the \emph{quench}, we
switch the field on abruptly at the start and hold it fixed.
This lets us test how the support grows as the system approaches
a mixed steady state.

\subsection{Models}
We use the dissipative transverse-field Ising
model~\cite{ates2012},
\begin{equation}
H(t)=J\sum_{\langle ij\rangle}Z_iZ_j+h(t)\sum_i X_i .
\end{equation}
Here $X_i$ and $Z_i$ are Pauli operators on site $i$,
$\langle ij\rangle$ denotes a nearest-neighbour bond, $J$ is the
Ising coupling, and $h(t)$ is the transverse field. We set $J=1$
and report time in units of $J^{-1}$, using $Jt$ on the plots.
The systems are open chains of 4--10 qubits and a $3\times3$
lattice. Each site has a jump $c_i=\sigma_i^-$ at rate $\gamma$,
with $\sigma^-=|1\rangle\langle0|$. The reference state is
the all-down state $|11\cdots1\rangle$.

The field jumps from zero to $h=0.5$ at $t=0$ in the quench.
We set $\gamma=1$ by default and also test $\gamma=0.5$
and $2$ in the cost--accuracy comparison. For the pulse,
$h(t)=0.75\sin^2(\pi t/2)$ for $0<t<2$ and $h(t)=0$ otherwise,
with $\gamma=1$.

The default circuit has four layers. Each layer contains
single-site rotations $R_x$ and $R_y$ generated by Pauli $X$
and $Y$, and a rotation $R_{zz}$ generated by $Z_iZ_j$ on each
bond. Each gate has its own angle. Santos \emph{et al.}~
\cite{santos2025} use only $R_x$ and $R_{zz}$. At zero angles,
the circuit is the identity and all the initial weight is on
the reference state.

\subsection{Fixed-rank baselines and threshold settings}
For the Hamming-ordered baseline, we keep the $R$ labels
closest to the reference bitstring in Hamming distance. To build
the hindsight baseline, we first run the adaptive method. We
then keep the reference bitstring and select the other labels
in order of their time-integrated weights from that run. We test
three hindsight support sizes per system, approximately $1/2$,
$3/4$, and $1$ times the peak adaptive $R$. Both baselines start
with all weight on the reference state and keep the same labels
throughout.

For all pulse sizes and the ten-qubit quench, we test four
thresholds on $\Gamma_{\rm leak}/\tau$, namely $2$, $3$, $5$
and $10$ in units of $10^{-3}$ per unit time, at $\kappa=0.3$.
At $3\times10^{-3}$ we also test $\kappa=0.2$ and $0.5$.
For the four-, six- and eight-qubit quenches, we test only
$3\times10^{-3}$ and $10^{-2}$, both at $\kappa=0.3$.

\subsection{Cost model}
A shot is one execution of a measurement circuit. We call
an individual overlap measurement used to build the
variational equations a \emph{primitive}. We count each
primitive circuit evaluation as one setting, regardless of how
many shots it uses. Let $N_\theta$ be the
number of gate angles and $n_P$ the number of Pauli operators
in the decomposition used to measure the Lindblad terms.
The derivative $\partial_k|\psi_a\rangle$ is with respect to
$\theta_k$, and $R_a$ counts the components with nonzero weight
at that evaluation.

We count $N_\theta R_a^2$ settings for the overlaps
$\langle\psi_a|\partial_k\psi_b\rangle$, and
$R_aN_\theta(N_\theta-1)/2$ for the real derivative overlaps
$\operatorname{Re}\langle\partial_k\psi_a|\partial_j\psi_a\rangle$.
For each Pauli operator $P$ in the decomposition, we also count
the state overlaps $\langle\psi_x|P|\psi_q\rangle$ and, across
all $P$, $2N_\theta n_P R_a^2$ settings for the derivative
overlaps $\langle\partial_k\psi_a|P|\psi_q\rangle$. The indices
$a,b,x,q$ denote basis labels. At checkpoints we add the costs
of the squared generator norm $\|\mathcal L[\rho]\|_2^2$,
candidate fluxes and the inflow guard whenever they are evaluated.

\subsection{Integration and regularisation}
We integrate with RK4 at time step $\Delta t=0.02$ and check
against $\Delta t=0.005$. We regularise the solve for the angle
block $M_{\theta\theta}$ with a smooth spectral
cutoff~\cite{medvidovic2023}. The absolute and relative cutoff
parameters are $a_c=r_c=10^{-6}$. A cutoff of $10^{-4}$ limits
accuracy. At $10^{-8}$, the condition number of the metric can
reach $10^{12}$ and the angles can jump along directions with
near-zero eigenvalues.

\subsection{Shot noise and certification batches}
For a real or imaginary overlap value $m\in[-1,1]$, we emulate
$N_i$ Hadamard-test shots~\cite{mitarai2019} by drawing a count
$K_i\sim\operatorname{Binomial}(N_i,(1+m)/2)$ and returning
$2K_i/N_i-1$. Shot noise can make the angle block indefinite.
In the robustness runs, let $\sigma_{\rm shot}=1/\sqrt{N_0}$,
where $N_0$ is the number of shots per measurement. This is an upper
bound on the standard error of a single Hadamard-test sample
mean. We use $\sigma_{\rm shot}$ to set three quantities. The leakage
threshold on $\Gamma_{\rm leak}/\tau$ is
$\max(10^{-2},5\sigma_{\rm shot})$. The removal threshold on
$\lambda_x$ is $\max(10^{-3},2\sigma_{\rm shot})$, and the
total removal budget is $\max(10^{-2},10\sigma_{\rm shot})$.
The balanced trigger keeps $\kappa=0.3$. A smooth cutoff
with $a_c=r_c=\sigma_{\rm shot}$ discards negative eigenvalues
and stabilises the solve. The separate weighted and uniform
allocation runs use the noise-free thresholds and
$a_c=r_c=10^{-3}$.

Let $E$ count the real-valued measurements needed for one
McLachlan evaluation, with real and imaginary parts counted
separately when both are needed. The nominal shot count $N_0$
is what each measurement would receive under uniform allocation.
The total budget for that evaluation is $B=EN_0$. The weighted
allocation reserves one
shot per measurement. It apportions the remaining $B-E$ in
proportion to nonnegative weights $w_i$, giving target counts
\begin{equation}
N_i=1+(B-E)\frac{w_i}{\sum_{j=1}^E w_j}.
\end{equation}
The weights are $\alpha_a\alpha_b$ for
$\langle\psi_a|\partial_k\psi_b\rangle$, $\alpha_a\alpha_q$
for $\langle\partial_k\psi_a|P|\psi_q\rangle$, $\alpha_a^2$
for $\operatorname{Re}\langle\partial_k\psi_a|\partial_j\psi_a\rangle$,
and $\alpha_x+\alpha_q$ for $\langle\psi_x|P|\psi_q\rangle$.
Appendix~\ref{app:d} relates these weights to the variance
of a linear estimate and to the purity of the state.
Related allocation schemes use Hamiltonian weights or metric
and gradient information in variational quantum
algorithms~\cite{wecker2015,arrasmith2020,vanstraaten2021,kubler2020,gu2021,zhang2026}.

Both schemes use the same per-amplitude sampling rule for
jump measurements. Their sampled support trajectories can
differ. The archived shot totals count McLachlan measurements.
The separate flux and inflow measurements are also counted in
the four-qubit comparison in Sec.~\ref{sec:weighted-results}.

For a pointwise certificate, fix the state, circuit and applied
velocities before collecting a fresh batch of paired outcomes.
This \emph{certification batch} is used to bound the residual
at that checkpoint (Appendix~\ref{app:b}). If checkpoint $m$ is assigned
failure probability $\delta_m$, choosing
$\sum_m\delta_m\leq\delta_{\rm tot}$ ensures joint checkpoint
coverage of at least $1-\delta_{\rm tot}$ by the union bound.
The finite-shot benchmarks use their shots for propagation
and support decisions.

\subsection{Reference and metrics}
We compute the reference trajectory by integrating the full
Lindblad equation with an adaptive eighth-order Runge--Kutta
method at relative tolerance $10^{-9}$, and check it against
the Quantum Toolbox in Python (QuTiP)~\cite{qutip1,qutip2}.
Mean trace distance (mean TD) averages the distance between
$\rho/\tau$ and $\rho^*$ over the reported time interval.
We use $F$ for the Uhlmann fidelity and $1-F$ for infidelity.

To see how many components suffice to capture most of the
exact state at a given time, we sort the eigenvalues of
$\rho^*(t)$ from largest to smallest. The \emph{oracle rank}
is the smallest number of eigenvalues whose sum reaches
$1-\epsilon$, with $\epsilon=10^{-3}$ in our benchmarks.
Thus the discarded eigenvalues carry at most $0.1\%$ of the
total weight. Here \emph{oracle} means access to the exact
state at each time. We use that information to benchmark
the support size $R$ in the pulse. The algorithm decides
whether to add states using the jump flux and residual
of its own variational state.

To compare costs, we first estimate how many circuits the
fixed-rank method needs to match the adaptive run's mean trace
distance. We interpolate between the two fixed-rank points that
bracket that error, using the logarithms of cost and error.
We divide the resulting circuit count by the adaptive count
to report the saving.

\section{Results}\label{sec:results}

\subsection{Numerical validation}\label{sec:validation}

\begin{figure*}[t]
\includegraphics[width=\textwidth]{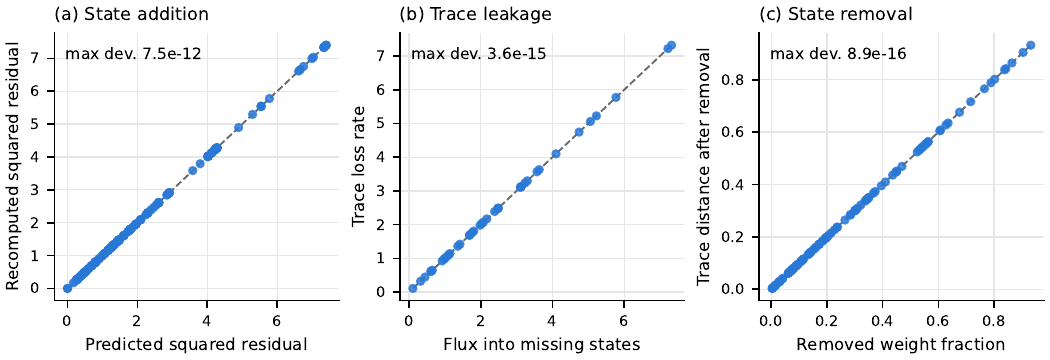}
\caption{Each point compares two calculations for the same
variational state. The dashed line is $y=x$ and each panel
reports the largest $|y-x|$. (a)~Predicted squared residual
$D_S^2-V_c^2$ (horizontal) and squared residual $D_{S\cup\{c\}}^2$
recomputed after adding $c$ at zero weight (vertical).
(b)~Sum of omitted-state fluxes $\sum_{c\notin S}V_c$
(horizontal) and trace loss from the retained weight
derivatives, $-\sum_{a\in S}\dot\alpha_a$ (vertical).
(c)~Normalised weight $\lambda_x$ of the deleted component
(horizontal) and trace distance between the normalised states
before and after removal (vertical).}
\label{fig:identities}
\end{figure*}

We checked the four rank identities on 40 random instances
with 2--5 qubits and decay, dephasing or thermal jumps, for a
total of 371 checks. Figure~\ref{fig:identities} shows the
addition, leakage and removal checks. The maximum absolute
deviation for the residual-splitting identity was
$8.9\times10^{-9}$.

For local decay, dephasing and thermal jumps, the weight
derivatives from Eq.~\eqref{eq:rate} agreed with those from the
McLachlan equations to within $1.3\times10^{-15}$ across 30
random instances with 2--5 qubits.

We also checked each measurement family against explicit Qiskit~\cite{qiskit2024} Hadamard-test circuits to within $10^{-10}$.

\subsection{Rank adaptation during a driven pulse}

\begin{figure*}[t]
\includegraphics[width=\textwidth]{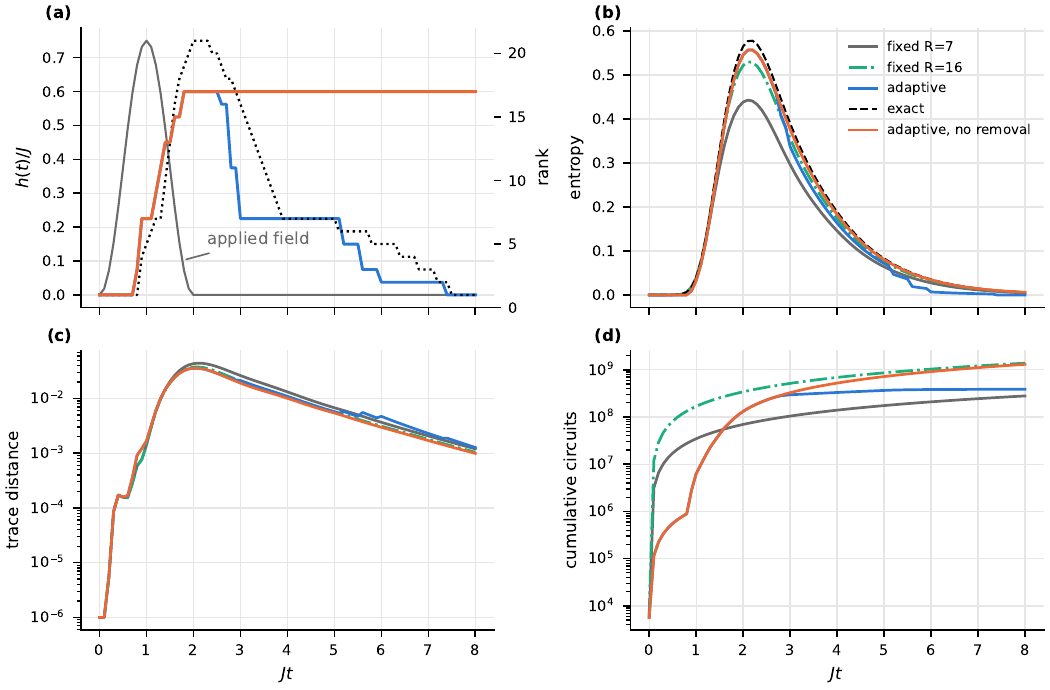}
\caption{Driven pulse for $n=6$ and $\gamma=J=1$.
(a)~Applied field $h(t)/J$ (gray, left axis) and support size
$R$ for the adaptive runs with and without removal (solid
curves, right axis). The dotted curve shows the oracle rank,
the fewest eigenvalues of the exact state that together account
for at least $99.9\%$ of its weight. (b)~von Neumann entropy, including the
exact reference. (c)~Trace distance to the exact state.
(d)~Cumulative measurement-circuit settings. Panels (b)--(d)
compare fixed $R=7$ and $R=16$ with the two adaptive runs.}
\label{fig:pulse}
\end{figure*}

By the time the field is turned off at $Jt=2$, the oracle rank
has peaked at 21 and both adaptive runs track 17 states
[Fig.~\ref{fig:pulse}(a)]. Their entropies rise during the
drive and fall during relaxation, following the trend of
the exact state [Fig.~\ref{fig:pulse}(b)].

After the drive, the run with removal is down to seven states
by $Jt=4$ and returns to $R=1$ by the end of the run. The run
without removal keeps all 17 states. Removal
reduces the total circuit count from $1.30\times10^9$ to
$3.87\times10^8$, with a small increase in mean trace distance
from $9.64\times10^{-3}$ to $1.03\times10^{-2}$. The cumulative
cost curves separate during relaxation as the support shrinks
[Fig.~\ref{fig:pulse}(d)].

Fixed $R=7$ uses fewer settings than the adaptive run with
removal and gives a larger mean trace distance. Fixed $R=16$
comes close to the adaptive error ($1.01\times10^{-2}$) and
uses about 3.6 times as many settings
[Fig.~\ref{fig:pulse}(c,d)].

\subsection{Scaling with system size}

\begin{figure*}[t]
\includegraphics[width=\textwidth]{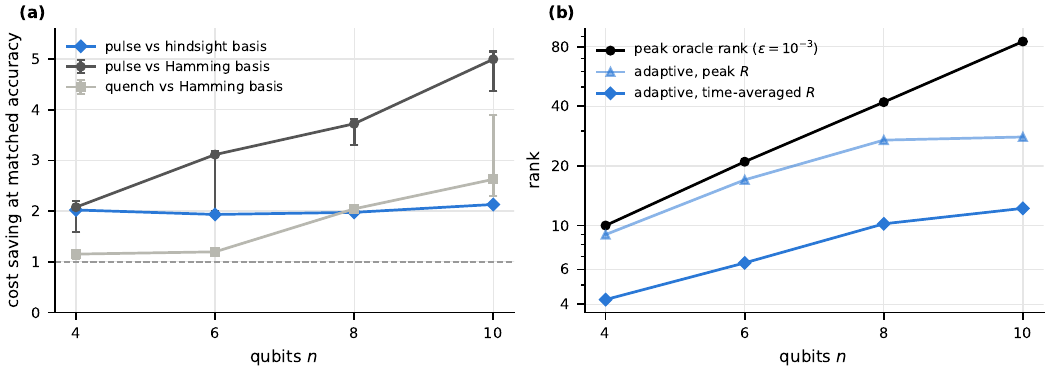}
\caption{Scaling with system size.
(a)~Fixed-rank circuit cost divided by adaptive cost at matched
mean trace distance within each system size.
For the Hamming basis, markers give the median and bars span the
minimum and maximum across threshold settings. The pulse uses six
settings at each size. The quench uses two at $n=4,6,8$ and six
at $n=10$. The hindsight-basis pulse comparison uses the default
setting at each size.
(b)~Peak oracle rank compared with the peak and time-averaged
adaptive support size $R$ during the pulse.}
\label{fig:scaling}
\end{figure*}

We vary the chain size from four to ten qubits while keeping
the circuit depth at four layers. Figure~\ref{fig:scaling}(a)
shows how the relative circuit cost changes with size.
Figure~\ref{fig:scaling}(b) compares the growth of the peak
oracle rank and the adaptive support during the pulse.

\emph{Growth in support size.} At the same discarded-weight
tolerance, the peak oracle rank of the pulse roughly doubles
with each pair of added sites, from 10 to 85 over the sizes
tested. The peak adaptive support grows from 9 to 28, and
its time average grows from 4.2 to 12.2
[Fig.~\ref{fig:scaling}(b) and Table~\ref{tab:scaling}].
The gap between the peak and the average reflects the
smaller support used during much of the trajectory.

\begin{table*}[t]
\caption{Pulse benchmark. Support sizes by system size for the
default adaptive trigger. The last column gives the range of
fixed ranks tested.}\label{tab:scaling}
\begin{ruledtabular}
\begin{tabular}{lllll}
$n$ & Peak oracle rank ($\epsilon  = 10^{-3}$) & Adaptive mean $R$ & Adaptive max $R$ & Fixed $R$ spanned \\
\hline
4 & 10 & 4.2 & 9 & 5--16 \\
6 & 21 & 6.5 & 17 & 1--32 \\
8 & 42 & 10.2 & 27 & 9--37 \\
10 & 85 & 12.2 & 28 & 11--60 \\
\end{tabular}
\end{ruledtabular}
\end{table*}

\begin{table*}[t]
\caption{Pulse benchmark at the default adaptive trigger against
the fixed-rank hindsight and Hamming-ordered baselines. Savings
are circuit-count ratios at matched mean trace distance within
each system size.}\label{tab:hindsight}
\begin{ruledtabular}
\begin{tabular}{lllll}
$n$ & Adaptive (mean TD, circuits) & Hindsight mean TD ($R$ shown) & Saving vs hindsight & Saving vs Hamming \\
\hline
4 & $7.6\times 10^{-3}$, $6.4\times 10^{7}$ & 1.15 / 0.82 / 0.70 $\times 10^{-2}$ ($R = 4$, 6, 9) & 2.0$\times$ & 2.1$\times$ \\
6 & $1.03\times 10^{-2}$, $3.9\times 10^{8}$ & 1.16 / 1.02 / 0.94 $\times 10^{-2}$ ($R = 8$, 12, 17) & 1.9$\times$ & 3.2$\times$ \\
8 & $1.25\times 10^{-2}$, $1.6\times 10^{9}$ & 1.36 / 1.23 / 1.16 $\times 10^{-2}$ ($R = 13$, 20, 27) & 2.0$\times$ & 3.8$\times$ \\
10 & $1.54\times 10^{-2}$, $3.5\times 10^{9}$ & 1.71 / 1.57 / 1.44 $\times 10^{-2}$ ($R = 14$, 21, 28) & 2.1$\times$ & 5.0$\times$ \\
\end{tabular}
\end{ruledtabular}
\end{table*}

\emph{Growth in circuit cost.} Figure~\ref{fig:scaling}(a)
compares costs within each system size. We use the mean trace
distance of the adaptive run as the target and interpolate
the fixed-rank cost at that accuracy
(Sec.~\ref{sec:benchmarks}). In the pulse, the median
fixed-to-adaptive circuit ratio for the Hamming basis rises
from 2.1 to 5.0. The bars show how much that ratio varies
with the trigger thresholds. The fixed hindsight basis uses about twice
as many settings as the adaptive run across the sizes tested,
even though it uses labels selected from the adaptive
trajectory (Table~\ref{tab:hindsight}). This comparison
shows the benefit of varying the support over time.

The adaptive pulse itself requires more work as the chain
grows. Its circuit count rises from $6.4\times10^7$ to
$3.5\times10^9$, and the mean trace distance rises from
$7.6\times10^{-3}$ to $1.54\times10^{-2}$
(Table~\ref{tab:hindsight}). The relative savings
describe the cost of reaching the accuracy attained at each size.

\subsection{Rank adaptation after a quench}

\begin{figure*}[t]
\includegraphics[width=\textwidth]{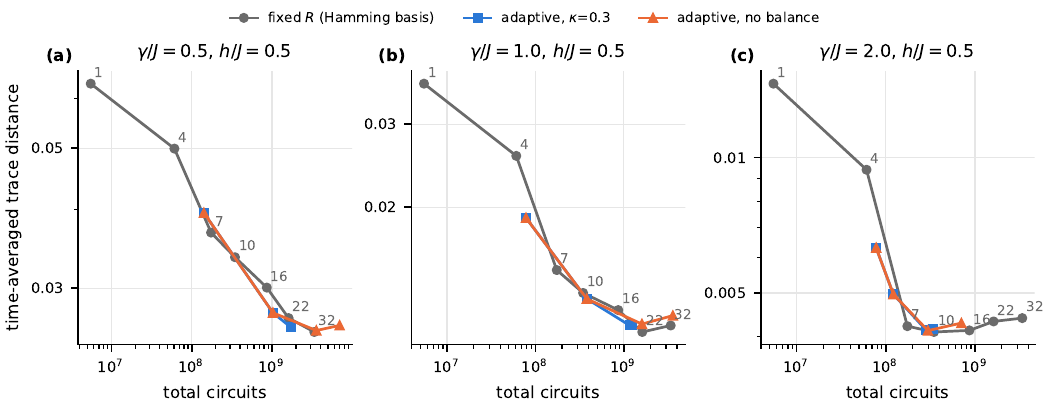}
\caption{Quench cost--accuracy comparisons at $n=6$ with the
post-quench field held at $h=0.5J$. The horizontal axis counts
circuit settings and the vertical axis gives mean trace distance.
Each fixed-rank point uses a Hamming-ordered basis and is
labelled by its support size $R$. Adaptive points use the
balanced trigger ($\kappa=0.3$) or the leakage threshold alone.
(a)~$\gamma=0.5J$, (b)~$\gamma=J$ and (c)~$\gamma=2J$.}
\label{fig:quench}
\end{figure*}

After the field is switched on, the entropy rises toward a
plateau and the support mainly grows. For the six-qubit chain,
the adaptive runs approach the best cost--accuracy trade-off
among the fixed-rank runs at the three dissipation rates in
Fig.~\ref{fig:quench}. At the strongest dissipation, a small
fixed basis consisting of the reference bitstring and its
one-bit neighbours already performs well.

The quench savings grow with system size
[Fig.~\ref{fig:scaling}(a)]. At ten qubits, the median
fixed-to-adaptive circuit ratio is 2.6 at matched accuracy.
The pulse offers more room for savings because the support
can also shrink during relaxation.

\subsection{\texorpdfstring{$3\times3$}{3 x 3} lattice and the removal budget}

On the $3\times3$ benchmark of Santos \emph{et al.}~
\cite{santos2025}, using their $R_xR_{zz}$ circuit, adaptive
rank reaches mean trace distance $9.3\times10^{-3}$ versus
$1.21\times10^{-2}$ at fixed $R=10$, using 13\% fewer circuits.

With the $R_xR_yR_{zz}$ circuit and the cumulative $10^{-2}$
removal budget, repeated deletions of components below
$10^{-3}$ reduce the final support to $R=1$. The corresponding
purity, $\operatorname{Tr}[(\rho/\tau)^2]$, is 1.000 versus
0.994 for the exact state. A tighter cumulative budget or a
budget applied over time windows retains more states
(Table~\ref{tab:budget}). For the pulse at $n=6$ and 8, the
windowed budget lowers the error by 3--4\% and increases
circuit counts by 32--42\%. The scaling runs use the
cumulative $10^{-2}$ budget.

\begin{table*}[t]
\caption{Effect of the removal budget on the $3\times3$
lattice with the $R_xR_yR_{zz}$ circuit.}\label{tab:budget}
\begin{ruledtabular}
\begin{tabular}{lllll}
Removal budget & Mean TD & Max $1-F$ & Circuits & Final $R$ \\
\hline
Fixed $R = 10$ (no removal) & $4.2\times 10^{-3}$ & $3.3\times 10^{-3}$ & $1.23\times 10^{9}$ & 10 \\
Cumulative $10^{-2}$ & $5.2\times 10^{-3}$ & $6.9\times 10^{-3}$ & $7.9\times 10^{8}$ & 1 \\
Cumulative $2\times 10^{-3}$ & $4.5\times 10^{-3}$ & $3.6\times 10^{-3}$ & $1.08\times 10^{9}$ & 7 \\
Windowed $2\times 10^{-3}$ per $2/J$ & $4.6\times 10^{-3}$ & $3.6\times 10^{-3}$ & $9.7\times 10^{8}$ & 4 \\
\end{tabular}
\end{ruledtabular}
\end{table*}

\subsection{Ablations}

\begin{figure*}[t]
\includegraphics[width=\textwidth]{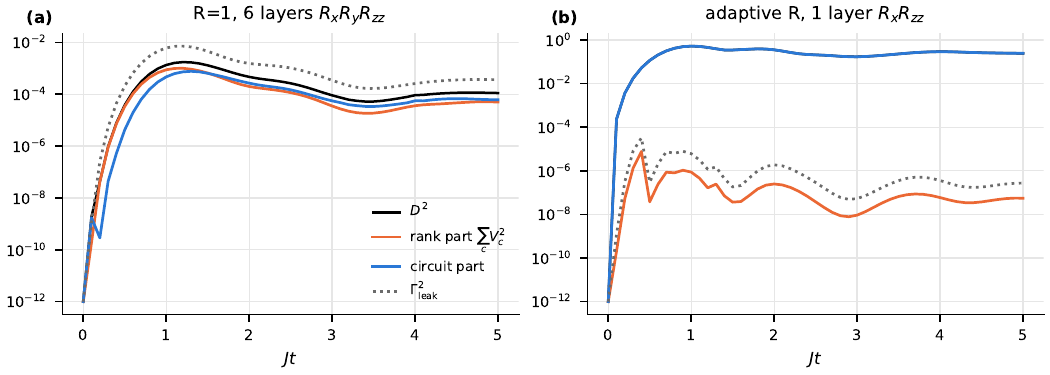}
\caption{Squared residual $D^2$, its rank contribution
$\sum_{c\notin S}V_c^2$, its circuit contribution
$D_{\rm circuit}^2$, and the leakage bound
$\Gamma_{\rm leak}^2$ in a six-qubit quench. (a)~Fixed $R=1$
and six $R_xR_yR_{zz}$ layers. (b)~Adaptive support and one
$R_xR_{zz}$ layer.}
\label{fig:splitting}
\end{figure*}
With $R=1$ and six circuit layers, restricting the support
accounts for about as much of the residual as restricting the
circuit. In the one-layer run, the circuit contribution is four
to six orders of magnitude larger than the rank contribution.
There is little to gain by adding states at the current angles
(Fig.~\ref{fig:splitting}). Turning off the balanced trigger
gives nearly the same mean trace distance, but uses 2.5 times
as many circuits for the pulse and 1.4 times as many for the
quench (Table~\ref{tab:ablation}).

The entropy-only trigger never adds a state. The leakage
trigger first fires at $Jt=0.4$ for the quench and $0.8$ for
the pulse. Searching only one-bit neighbours or sampling jumps
gives the same support decisions as exhaustive screening in
these runs. Disabling the inflow guard increases quench
removals from 6 to 22, with states removed and re-added.

\begin{table*}[t]
\caption{Ablations at $n=6$ (base thresholds $3\times10^{-3}$ and $\kappa=0.3$). Circuits are totals over the run.}\label{tab:ablation}
\begin{ruledtabular}
\begin{tabular}{lllll}
Variant & Pulse mean TD & Pulse circuits & Quench mean TD & Quench circuits \\
\hline
Full protocol & $1.03\times 10^{-2}$ & $3.9\times 10^{8}$ & $1.12\times 10^{-2}$ & $1.17\times 10^{9}$ \\
Entropy-only trigger & $3.35\times 10^{-2}$ & $8.8\times 10^{6}$ & $3.66\times 10^{-2}$ & $5.5\times 10^{6}$ \\
No removal & $9.6\times 10^{-3}$ & $1.30\times 10^{9}$ & $1.12\times 10^{-2}$ & $1.23\times 10^{9}$ \\
No balance ($\kappa$ off) & $9.6\times 10^{-3}$ & $9.6\times 10^{8}$ & $1.13\times 10^{-2}$ & $1.60\times 10^{9}$ \\
No inflow guard & $1.04\times 10^{-2}$ & $3.5\times 10^{8}$ & $1.24\times 10^{-2}$ & $6.0\times 10^{8}$ \\
No entropy scheduler & $1.15\times 10^{-2}$ & $2.8\times 10^{8}$ & $1.17\times 10^{-2}$ & $9.8\times 10^{8}$ \\
Exact discovery (all $2^{n}$ states) & $1.03\times 10^{-2}$ & $3.9\times 10^{8}$ & $1.12\times 10^{-2}$ & $1.17\times 10^{9}$ \\
Jump-sampling discovery (4000 shots) & $1.03\times 10^{-2}$ & $3.9\times 10^{8}$ & $1.12\times 10^{-2}$ & $1.17\times 10^{9}$ \\
\end{tabular}
\end{ruledtabular}
\end{table*}

\subsection{Numerical evaluation of the continuous-time bound}

\begin{figure*}[t]
\includegraphics[width=\textwidth]{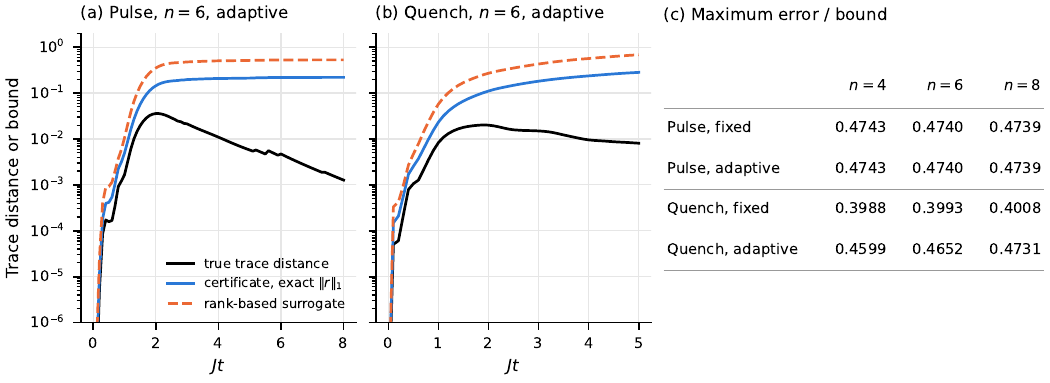}
\caption{Numerical evaluation of the continuous-time
trace-distance bound. (a,b)~Six-qubit adaptive pulse and
quench trajectories. The curves show the trace distance,
the bound integrated with the exact residual trace norm
$\|r\|_1$, and a surrogate bound that integrates the upper
estimate $\sqrt{\operatorname{rank}(r)}D_{\rm app}$ in its place.
(c)~For each of the
12 runs, the maximum of the trace distance divided by the bound
integrated with $\|r\|_1$, over recorded times $t>0$. A ratio
above 1 would indicate that the numerical bound fell below
the trace distance. At $t=0$ both are zero. The fixed runs
use the Hamming-ordered basis with $R=n+1$.}
\label{fig:certificate}
\end{figure*}

Across all 792 recorded times in the 12 runs at $n=4,6,8$,
the measured trace distance was at or below the bound integrated
with $\|r\|_1$. At $t>0$, the largest ratio of trace distance to
bound was about 0.47. It occurred in the four-qubit fixed-rank
pulse at $Jt=0.3$ [Fig.~\ref{fig:certificate}(c)].

To see how far above the error the bounds lie, we divide each
bound by the measured trace distance. We use the times when
that distance exceeds $10^{-4}$.
The per-run medians of this ratio range from 6.3 to 22.5 for
the bound from $\|r\|_1$ and from 9.3 to 86.4 for the
simulated-rank surrogate. In the six-qubit pulse, the bound
levels off while dissipation reduces the error. The surrogate
bound itself exceeds 1 in three eight-qubit runs (adaptive
pulse, fixed quench and adaptive quench). The structural
bound $\nu$ can replace the simulated rank for measurement-based
evaluation, at the cost of a potentially larger bound.

\subsection{Finite shots}

Finite sampling can affect both the variational updates and the
decisions to add or remove states. To test these effects, we
emulate Hadamard-test statistics for every primitive in four-qubit
runs with three seeds. Adaptive rank and fixed $R=5$ both give a mean trace
distance of $4.5\times10^{-2}$ at $10^5$ shots per
measurement. The fixed-rank run uses 1.3 times as many circuit
settings. At $2\times10^3$ to $10^4$ shots, the noise-scaled
thresholds keep the support between 1 and 3. The error is
about 10\% higher, with circuit counts lower by factors of
7--20. The regulariser accounts for most of the error at
these budgets ($2.8\times10^{-2}$ without shot noise).

\subsection{Weighted shot allocation}\label{sec:weighted-results}

\begin{figure*}[t]
\includegraphics[width=\textwidth]{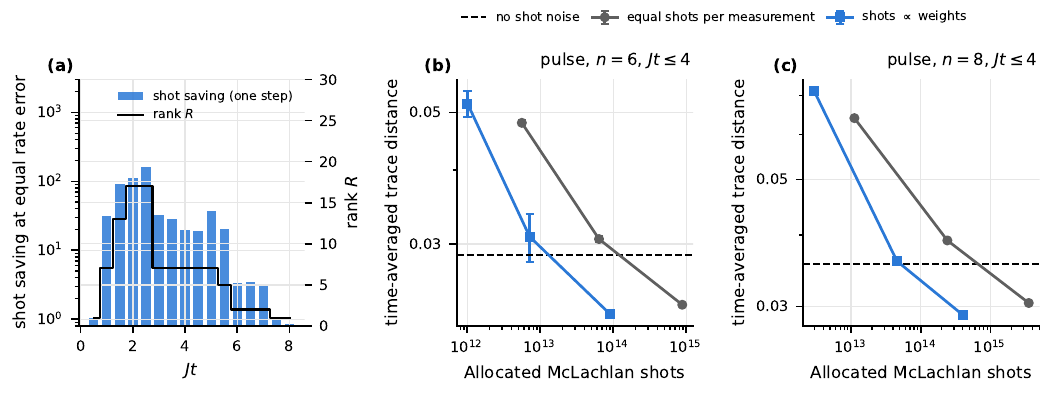}
\caption{Weighted and uniform shot allocation for the driven
pulse. (a)~Ratio of uniform to weighted shots per McLachlan
evaluation at equal error in $\dot\rho$ along the six-qubit
trajectory (bars), with support size $R$ (line).
(b,c)~Mean trace distance through $Jt=4$ versus total
allocated McLachlan shots. From left to right, blue squares
give weighted budgets of $10^3$, $10^4$ and $10^5$ nominal
shots per measurement. Gray circles give uniform budgets of
$10^4$, $10^5$ and $10^6$. Panel (b)~averages over two seeds
at six qubits, with bars spanning the seed results.
Panel (c)~uses seed 0 at eight qubits. Dashed lines give the
corresponding noise-free errors.}
\label{fig:shots}
\end{figure*}

\emph{One step.} In the six-qubit pulse, we compare the shot
counts needed for equal error in $\dot\rho$ over 20 noise
repetitions. The ratio of uniform to weighted shots is
22--33 at $R=7$, 103--155 at $R=17$, and 38 when summed
over the trajectory [Fig.~\ref{fig:shots}(a)]. At $R=1$
the allocation gives no saving.

\emph{Whole runs.} We compare total allocated McLachlan shots
through $Jt=4$ [Fig.~\ref{fig:shots}(b,c) and
Table~\ref{tab:shots}]. We save fewer shots over a whole run because shot noise accounts
for only part of the trajectory error. A fluctuation can also
cause the controller to add a state. In some runs, this gives
a smaller error than in the noise-free run.

At $10^3$ nominal shots per measurement, the weighted scheme
gives mean trace distances of 0.05175 at six qubits and
0.07146 at eight. In both cases the lowest tested uniform
budget already gives a smaller error, so the saving is
bounded above. At $10^4$ nominal shots, the interpolated
uniform total is 8.5 times the weighted total for six
qubits and 13 times for the single-seed eight-qubit run.
Matching both schemes at two targets relative to the
noise-free error gives factors of 5.0 and 6.6 at six qubits
(Appendix~\ref{app:d}).

We also repeated the four-qubit pulse through $Jt=4$ with
separate counters for variational and screening shots and two
seeds per budget. Weighted allocation at $10^4$ nominal shots
per measurement used $8.78\times10^{11}$ shots including
screening and reached mean trace distance 0.0214. Uniform
allocation at $10^5$ used $9.63\times10^{12}$ shots and
reached 0.0217. These counts are from classical emulation.

\begin{table*}[t]
\caption{Allocated McLachlan shots for each weighted point in
Fig.~\ref{fig:shots}(b,c), through $Jt=4$. The six-qubit values
average seeds 0 and 1. The eight-qubit values use seed 0.
Each row uses the measured weighted mean trace distance (TD) as
its target. We interpolate the uniform cost where the tested
budgets bracket the target. If the lowest tested uniform budget
already reaches the target, it gives an upper bound ($\leq$).
If the highest budget misses the target, it gives a lower bound
($>$) over the tested cost range. The saving is the uniform cost
divided by the weighted cost.}
\label{tab:shots}
\begin{ruledtabular}
\begin{tabular}{rrrrrr}
$n$ & Nominal weighted budget & Mean TD & Weighted shots
    & Uniform shots at target & Saving \\
\hline
6 & $10^3$ & 0.05175 & $1.01\times10^{12}$
  & $\leq 5.60\times10^{12}$ & $\leq 5.6\times$ \\
6 & $10^4$ & 0.03086 & $7.17\times10^{12}$
  & $6.08\times10^{13}$ & $8.5\times$ \\
6 & $10^5$ & 0.02284 & $9.10\times10^{13}$
  & $>8.82\times10^{14}$ & $>9.7\times$ \\
8 & $10^3$ & 0.07146 & $2.95\times10^{12}$
  & $\leq 1.121\times10^{13}$ & $\leq 3.8\times$ \\
8 & $10^4$ & 0.03598 & $4.58\times10^{13}$
  & $5.89\times10^{14}$ & $13\times$ \\
8 & $10^5$ & 0.02901 & $3.98\times10^{14}$
  & $>3.54\times10^{15}$ & $>8.89\times$ \\
\end{tabular}
\end{ruledtabular}
\end{table*}

\section{Discussion}


The residual split tells us where to look when the
variational fit is poor. If much of the residual comes from
omitted states, there is room to improve the fit by keeping
more of them. If the circuit accounts for most of the
residual, a more expressive circuit may help. One way to
build it is to add generators with an operator-adaptive
method~\cite{chen2024}.

The pulse is where we gain the most from removing states.
Once the drive is off, the chain relaxes and the simulation
can let go of states it needed during the pulse. The
hindsight baseline already knows which labels carry the most
weight over the run. The saving against that baseline shows
what can be gained by changing how many states are kept and
when. In the quench, the support mainly grows. Here the
method can approach the fixed-rank cost--accuracy trade-off
without trying several values of $R$ first.

We designed the method for dynamics whose effective rank
stays well below $2^n$. Our benchmarks emulate the
measurements on a classical computer and compare with exact
reference trajectories up to ten qubits. Circuit counts
follow the model in Sec.~\ref{sec:benchmarks}. Jump sampling
can find candidates without searching the full basis. How
many measurements it needs on larger devices, and how well
it tolerates noise there, remain open questions. On a small
device, one could check whether the state with the largest
measured flux also gives the largest drop in squared
residual, then use that measurement to choose a state during
evolution.

In the benchmarks, we integrate the residual numerically.
To turn checkpoint measurements of $D_{\rm app}$ into a
hardware certificate, we would need to account for sampling
uncertainty, bound the residual between checkpoints, and
control the error from finite time steps. The mixture
weights also help us decide how to distribute the shots
used to estimate the variational equations.

Rank adaptation could also be combined with error
mitigation~\cite{cai2023}, projected variational
dynamics~\cite{barison2021}, or formulations that avoid
measuring the geometric tensor~\cite{gacon2024}.
\section*{Data availability}
The figures and tables report the numerical results of this study.

\section*{Code availability}
The code used for the numerical experiments is available at
\url{https://github.com/plain12356/adaptive-rank-variational-simulation}.

\begin{acknowledgments}
This work was supported by the National Science and Technology Council (Taiwan) under grant NSTC 114-2112-M-182-002-MY3 and by Chang Gung Memorial Hospital under grant BMRPL94.

\end{acknowledgments}

\section*{Author contributions}
Y.-H.C. conceived the topic, wrote the initial draft, and performed numerical experiments. R.W. supervised the project, drafted and reviewed the manuscript.

\section*{Competing interests}
The authors declare no competing interests.

\appendix

\section{Derivations of the rank identities}\label{app:a}

We work in the complete rotated basis
$|\psi_x\rangle=U(\theta)|x\rangle$, with projectors $P_x$ as in
Sec.~\ref{sec:ansatz}. Write $D^2(\dot\beta)$ for the squared
Hilbert--Schmidt residual at an arbitrary velocity $\dot\beta$.
Expanding that squared norm gives
\begin{align}
D^2(\dot\beta)&=\|\mathcal L[\rho]\|_2^2-2V^T\dot\beta
                 +\dot\beta^TM\dot\beta,\nonumber\\
D^2&=\|\mathcal L[\rho]\|_2^2-V^TM^{+}V
     \leq D^2(\dot\beta).
\end{align}
Here $M^{+}$ is the Moore--Penrose pseudoinverse of $M$.

\emph{Weight block.} The weight derivative is
$\partial\rho/\partial\alpha_x=P_x$. For a gate angle,
$\partial_kP_a=|\partial_k\psi_a\rangle\langle\psi_a|$ plus
its adjoint. Thus
\begin{align}
M_{\alpha_x\alpha_y}&=\operatorname{Tr}(P_xP_y)=\delta_{xy},
\nonumber\\
M_{\alpha_x\theta_k}
&=2\sum_{a\in S}\alpha_a\operatorname{Re}
 \big[\langle\psi_x|\partial_k\psi_a\rangle
       \langle\psi_a|\psi_x\rangle\big]\nonumber\\
&=\alpha_x\partial_k\langle\psi_x|\psi_x\rangle=0 .
\end{align}
The Kronecker delta $\delta_{xy}$ equals one for $x=y$
and zero otherwise. The two blocks decouple and
$\dot\alpha_x=V_x=\langle\psi_x|\mathcal L[\rho]|\psi_x\rangle$,
where $V_x$ denotes the weight entry of $V$.

\emph{State addition.} With $\alpha_c=0$ for $c\notin S$,
the state and its angle derivatives are unchanged, so
$M_{\theta\theta}$ and $V_\theta$ stay the same. The weight
block gains one unit diagonal entry and $D^2$ decreases
by $V_c^2$. Since $\rho|\psi_c\rangle=0$, the commutator
and anticommutator contribute zero to $V_c$, leaving
\begin{align}
V_c&=\sum_{k=1}^K\gamma_k
  \langle\psi_c|c_k\rho c_k^\dagger|\psi_c\rangle\nonumber\\
&=\sum_{k=1}^K\gamma_k\sum_{a\in S}
  \alpha_a|\langle\psi_c|c_k|\psi_a\rangle|^2\geq0 .
\end{align}

\emph{Trace leakage.} The gate-angle velocities preserve the
trace because $\operatorname{Tr}(\partial_kP_x)=0$. With
unconstrained weight velocities,
$\dot\tau=\sum_{x\in S}\dot\alpha_x$. The generator satisfies
$\operatorname{Tr}\mathcal L[\rho]=0$, so summing its diagonal
entries over the complete rotated basis gives
\begin{equation}
0=\sum_{x\in S}V_x+\sum_{c\notin S}V_c,
\qquad -\dot\tau=\sum_{c\notin S}V_c .
\end{equation}

\emph{Residual splitting.} Adding a set $C$ of omitted
labels at zero weight adds mutually orthogonal weight
directions. Thus
$D_{S\cup C}^2=D_S^2-\sum_{c\in C}V_c^2$. Taking $C$ to be
the full complement of $S$ gives the circuit residual.
Nonnegative $V_c$ yield
$\sum_{c\notin S}V_c^2\leq(\sum_{c\notin S}V_c)^2
=\Gamma_{\rm leak}^2$.

\emph{Rate equation.} The diagonal of $[H,\rho]$ vanishes
in the rotated basis because $\rho$ is diagonal there.
Completeness gives
$\langle\psi_x|c_k^\dagger c_k|\psi_x\rangle
=\sum_y|\langle\psi_y|c_k|\psi_x\rangle|^2$.
Consequently, the two dissipative terms in the weight
derivative are
\begin{align}
\dot\alpha_x
&=\sum_{k=1}^K\gamma_k\left(
   \sum_{a\in S}\alpha_a|\langle\psi_x|c_k|\psi_a\rangle|^2
   \right.\nonumber\\
&\hphantom{=\sum_{k=1}^K\gamma_k\big(}\left.
   {}-\alpha_x\langle\psi_x|c_k^\dagger c_k|\psi_x\rangle
   \right)\nonumber\\
&=\sum_{a\in S}W_{a\to x}\alpha_a
   -\alpha_x\sum_y W_{x\to y}.
\end{align}
The outflow to labels $y\notin S$ is $\Gamma_{\rm leak}$.
Each contribution to $W_{a\to x}$ is the jump rate
$\gamma_k\|c_k|\psi_a\rangle\|^2$ times the conditional
probability $p(x\mid a,k)$ of measuring $x$ after the
jump and $U^\dagger$.

\emph{State removal.} For a deleted label $x$ with
$\lambda_x<1$, the eigenvalues associated with $y\ne x$
increase from $\lambda_y$ to $\lambda_y/(1-\lambda_x)$.
The eigenvalue for $x$ changes from $\lambda_x$ to zero.
Using the common eigenbasis,
\begin{align}
\frac12\left\|\frac{\rho}{\tau}-
      \frac{\rho-\alpha_xP_x}{\tau-\alpha_x}\right\|_1
&=\frac12\left[\lambda_x+
  \sum_{y\ne x}\frac{\lambda_y\lambda_x}{1-\lambda_x}\right]
  \nonumber\\
&=\lambda_x .
\end{align}
\section{The trace-norm certificate}\label{app:b}

\emph{Proof of the theorem.} Let $e(t)=\rho(t)-\rho^*(t)$.
On an interval without updates, $\dot e=\mathcal L_t[e]+r$.
For $s_0\leq t$ in this interval, Duhamel's formula gives
\begin{equation}
e(t)=\mathcal P(t,s_0)e(s_0)+
      \int_{s_0}^t\mathcal P(t,s)r(s)\,ds,
\end{equation}
where $\mathcal P(t,s)$ is the propagator from $s$ to $t$.
At an interval endpoint, use the corresponding one-sided
state. The propagator is completely positive and trace
preserving. Positive trace-preserving maps contract the
trace norm of Hermitian operators~\cite{ruskai1994}, so
$\|e(t)\|_1\leq\|e(s_0)\|_1+\int_{s_0}^t\|r(s)\|_1\,ds$.
At an update, $e(t_j^+)=e(t_j^-)+\Delta_j$, and the
triangle inequality bounds the increase in the norm by
$\|\Delta_j\|_1$. Iterating from $e(0)=0$ gives
$\|\rho(t)-\rho^*(t)\|_1\leq C(t)$, where $C(t)$ is the
integral plus the update sum in Eq.~\eqref{eq:certificate}.

Since $\operatorname{Tr}\rho^*=1$,
$|1-\tau|\leq\|\rho-\rho^*\|_1\leq C(t)$. Now write
\begin{equation}
\frac{\rho}{\tau}-\rho^*
=\frac{\rho-\rho^*}{\tau}+\rho^*\,\frac{1-\tau}{\tau}.
\end{equation}
Using $\|\rho^*\|_1=1$ and dividing the trace norm by two
gives $\mathrm{TD}(\rho/\tau,\rho^*)\leq C(t)/\tau$.
$\square$

\emph{Residual factorisation.} The generator is
$-i(H_{\rm eff}\rho-\rho H_{\rm eff}^\dagger)
 +\sum_{k=1}^K\gamma_k c_k\rho c_k^\dagger$.
In $r$, the weight derivatives multiply $P_a$, the angle
derivatives pair $|\dot\psi_a\rangle$ with $|\psi_a\rangle$,
and the effective Hamiltonian pairs $H_{\rm eff}|\psi_a\rangle$
with $|\psi_a\rangle$. Each jump contributes the outer product
of $c_k|\psi_a\rangle$ with itself, multiplied
by $-\gamma_k\alpha_a$. These outer products and their
coefficients give $r=YGY^\dagger$ and the column count in
Eq.~\eqref{eq:rankresbound}.

\emph{Numerical evaluation.} The certificate already uses
the regularised velocities along the continuous trajectory.
In the numerical runs, we also have to account for time
stepping and quadrature.

\begin{enumerate}
\item \textbf{Time stepping.} Intermediate RK4 stage clipping
changes the numerical integrator and contributes to its
time-stepping error. Accepted endpoint clipping contributes
to the discrete update sum. At fixed rank, refining
$\Delta t=0.02$ to $0.005$ gives a trace distance between the
two trajectories at $Jt=4$ of $2.5\times10^{-6}$ for $n=4$
and $5.3\times10^{-6}$ for $n=6$. Both are small compared
with the bounds of order $10^{-1}$.

\item \textbf{Quadrature.} We use the trapezoidal rule on the
time grid for $\int_0^t\|r(s)\|_1\,ds$. It has error of order
$(\Delta t)^2$ where the integrand is smooth. A rigorous
upper bound also needs control between sampling times. In
our checks a left Riemann sum undercounted when the
residual grew from zero.
\end{enumerate}
\emph{Estimating the applied residual with shots.}
Fix the state, circuit and applied velocities before
sampling. Write
\begin{equation}
r=\sum_j b_j|u_j\rangle\langle v_j|,
\end{equation}
where $b_j$ is a scalar and $|u_j\rangle$ and $|v_j\rangle$
are unit vectors. The weight derivative contributes
$\dot\alpha_aP_a$. For a Pauli rotation,
$|d_{ka}\rangle=2i\partial_k|\psi_a\rangle$ is a unit vector
prepared by the insertion of that gate's Pauli generator.
The angle derivative contributes
$-i\alpha_a\dot\theta_k|d_{ka}\rangle\langle\psi_a|/2$
and its adjoint. A generator term $\xi P\rho Q$, with scalar
coefficient $\xi$ and Hermitian Pauli operators $P,Q$,
contributes $-\xi\alpha_a|P\psi_a\rangle\langle Q\psi_a|$
to $r$ for each component $a$.

Expanding the squared Hilbert--Schmidt norm gives
\begin{equation}
D_{\rm app}^2=\sum_{ij}b_i^*b_j
  \langle u_i|u_j\rangle\langle v_j|v_i\rangle .
\end{equation}
If $\Lambda_r=\sum_j|b_j|=0$, then $r=0$. Otherwise draw
$i,j$ independently with probabilities $p_\ell=|b_\ell|/\Lambda_r$.
For the drawn indices, set $\phi_\ell=\arg b_\ell$ and define
\begin{align}
A_{ij}&=e^{i(\phi_j-\phi_i)}\langle u_i|u_j\rangle,\nonumber\\
B_{ij}&=\langle v_j|v_i\rangle.
\end{align}
For each pair, we choose with equal probability to measure
both real parts or both imaginary parts. We perform a
separate Hadamard test on each overlap and include the known
phase in the first test. The outcomes are $Z_1,Z_2\in\{-1,1\}$.
We record $Z=Z_1Z_2$ for the real parts and $Z=-Z_1Z_2$ for
the imaginary parts. The tests are independent, so $\mathbb E[Z\mid i,j]=\operatorname{Re}(A_{ij}B_{ij})/2$.
Averaging over the index draws yields
$D_{\rm app}^2=2\Lambda_r^2\mathbb E[Z]$. Each of the $N$
pairs uses two fresh shots, for $2N$ shots in total. The
two executions remain independent when $i=j$. Reusing
propagation estimates can bias the product.

Taking the square root of the nonnegative part of
Eq.~\eqref{eq:ducb} gives the upper confidence bound (UCB)
\begin{equation}
D_{\rm UCB}=\sqrt{\max\!\left\{0,\widehat d+
  2\Lambda_r^2\sqrt{\frac{2\ln(2/\delta)}{N}}\right\}} .
\end{equation}
Then $\sqrt{\nu}\,D_{\rm UCB}$ bounds the residual trace
norm at the checkpoint with the same confidence. At fixed
sample size, a larger $\Lambda_r$ increases the statistical
margin.

\section{Hilbert--Schmidt non-contractivity}\label{app:c}

Let $X(t)$ be a traceless Hermitian operator evolving under
a time-independent Lindblad generator. Its squared
Hilbert--Schmidt norm changes according to
\begin{align}
\frac{d}{dt}\|X\|_2^2
&=2\operatorname{Tr}(X\mathcal L[X])
 =2\langle X,\mathcal L_s[X]\rangle_{\rm HS},\nonumber\\
\mathcal L_s&=\tfrac12(\mathcal L+\mathcal L^\dagger).
\end{align}
Here $\langle A,B\rangle_{\rm HS}=\operatorname{Tr}(A^\dagger B)$,
and $\mathcal L^\dagger$ is the adjoint in this inner product.
Let $\mu$ be the largest eigenvalue of the compression of
$\mathcal L_s$ to traceless Hermitian operators. Equivalently,
$\mu$ is the maximum of $\langle X,\mathcal L_s[X]\rangle_{\rm HS}$
over unit-norm traceless Hermitian $X$. If $\mu\leq0$, the
Hilbert--Schmidt distance between any two evolving states
is nonincreasing. If $\mu>0$, take a maximiser $X_{\max}$
and initial states $I/2^n\pm\varepsilon X_{\max}$ with
$\varepsilon>0$ small enough for positivity. Their
Hilbert--Schmidt distance has initial relative growth rate
$\mu$.

The Hamiltonian term $-i[H,\cdot]$ is skew-adjoint in this
inner product and cancels in $\mathcal L_s$. For
$\sigma^-$ decay at $\gamma=1$, the values of $\mu$ for
$n=1$ through 5 are $-0.50$, $-0.29$, $-0.09$, $+0.12$
and $+0.33$, independent of $J$ and $h$. At $n=4$, the
pair above has Hilbert--Schmidt distance increasing from
0.1102 to 0.1114 over $Jt=0.1$. Its trace distance falls
from 0.171 to 0.162. The contractivity argument based on
the Hilbert--Schmidt norm fails for the four- and
five-qubit examples. The trace-norm proof uses
contractivity of the positive trace-preserving propagator.

\begin{table}[t]
\caption{Allocated McLachlan shots for the six-qubit pulse at
matched accuracy. Each measured budget is averaged over two
seeds. The ratio is uniform shots divided by weighted shots.}
\label{tab:fixedacc}
\begin{ruledtabular}
\begin{tabular}{@{}ccccc@{}}
Target factor & Mean TD & Uniform & Weighted & Ratio \\
\hline
1.25 & 0.0360 & $2.65\times10^{13}$ & $3.99\times10^{12}$ & 6.6 \\
1.50 & 0.0432 & $9.89\times10^{12}$ & $2.00\times10^{12}$ & 5.0 \\
\end{tabular}
\end{ruledtabular}
\end{table}

\section{Weight-proportional shot allocation}\label{app:d}

\emph{Variance of a linear estimate.} Let
$f=\sum_{i=1}^E w_i x_i$ be a linear combination with
nonnegative coefficients $w_i$. The estimate $\widehat x_i$
of measurement $x_i$ uses $N_i$ shots and has variance
$\sigma_i^2/N_i$, where $\sigma_i^2\leq1$ is its single-shot
variance. For independent estimates and total budget
$B=\sum_iN_i$, the estimate $\widehat f=\sum_i w_i\widehat x_i$
has variance
\begin{equation}
\operatorname{Var}(\widehat f)=\sum_i\frac{w_i^2\sigma_i^2}{N_i}.
\end{equation}
Allowing continuous allocations and ignoring the one-shot
minimum, the optimum assigns $N_i\propto w_i\sigma_i$.
The optimal and uniform variances are
\begin{align}
\operatorname{Var}_{\rm opt}(\widehat f)
&=\frac{(\sum_i w_i\sigma_i)^2}{B},\nonumber\\
\operatorname{Var}_{\rm uniform}(\widehat f)
&=\frac{E\sum_i w_i^2\sigma_i^2}{B}.
\end{align}
When the variance is nonzero, the ratio
$G_{\rm shot}=\operatorname{Var}_{\rm uniform}(\widehat f)/
\operatorname{Var}_{\rm opt}(\widehat f)$ is at least one by
Cauchy--Schwarz. With equal single-shot variances the optimum
reduces to allocating in proportion to $w_i$.

For a family with all $R_a^2$ active pairs and weights
$w_{ab}=\alpha_a\alpha_b$, assume equal single-shot
variances. Write $p_2=\sum_a\lambda_a^2$ for the purity
of the normalised state and $R_{\rm eff}=1/p_2$ for the
participation ratio of its weights. This is the number of
equally weighted components that would give the same purity.
Then
\begin{equation}
G_{\rm shot}=R_a^2\frac{(\sum_a\alpha_a^2)^2}{(\sum_a\alpha_a)^4}
   =(R_a p_2)^2=\left(\frac{R_a}{R_{\rm eff}}\right)^2 .
\end{equation}
The more unevenly the weights are distributed, the more this
allocation helps. The angle update depends nonlinearly on
the estimates of $M_{\theta\theta}$ and $V_\theta$. We can
linearise the update to estimate the error to first order.
How strongly each measurement affects that error also
depends on the conditioning of the metric. Summing the
predictions across measurement families gives gains of 30
at $R=7$ and 122 at $R=17$. The measured ranges are 22--33
and 103--155, respectively.

\emph{Whole-trajectory saving at matched accuracy.} The
noise-free mean trace distance for the six-qubit pulse
through $Jt=4$ is 0.02880. We compare the two seed-averaged
curves in Fig.~\ref{fig:shots}(b) at 1.25 and 1.50 times
that error. For each scheme, we interpolate linearly in
log shots and log mean trace distance within the tested
budget range (Table~\ref{tab:fixedacc}).

\bibliography{references}

@book{breuer2007,
  author = {Breuer, H.-P. and Petruccione, F.},
  title = {The Theory of Open Quantum Systems},
  publisher = {Oxford University Press},
  address = {Oxford},
  year = {2007},
  isbn = {9780199213900}
}

@article{lindblad1976,
  author = {Lindblad, G.},
  title = {On the generators of quantum dynamical semigroups},
  journal = {Commun. Math. Phys.},
  volume = {48},
  pages = {119--130},
  year = {1976}
}

@article{gks1976,
  author = {Gorini, V. and Kossakowski, A. and Sudarshan, E. C. G.},
  title = {Completely positive dynamical semigroups of {N}-level systems},
  journal = {J. Math. Phys.},
  volume = {17},
  pages = {821--825},
  year = {1976}
}

@article{weimer2021,
  author = {Weimer, H. and Kshetrimayum, A. and Or{\'u}s, R.},
  title = {Simulation methods for open quantum many-body systems},
  journal = {Rev. Mod. Phys.},
  volume = {93},
  pages = {015008},
  year = {2021}
}

@article{mclachlan1964,
  author = {McLachlan, A. D.},
  title = {A variational solution of the time-dependent {S}chrodinger equation},
  journal = {Mol. Phys.},
  volume = {8},
  pages = {39--44},
  year = {1964}
}

@article{yuan2019,
  author = {Yuan, X. and Endo, S. and Zhao, Q. and Li, Y. and Benjamin, S. C.},
  title = {Theory of variational quantum simulation},
  journal = {Quantum},
  volume = {3},
  pages = {191},
  year = {2019}
}

@article{endo2020,
  author = {Endo, S. and Sun, J. and Li, Y. and Benjamin, S. C. and Yuan, X.},
  title = {Variational quantum simulation of general processes},
  journal = {Phys. Rev. Lett.},
  volume = {125},
  pages = {010501},
  year = {2020}
}

@article{yoshioka2020,
  author = {Yoshioka, N. and Nakagawa, Y. O. and Mitarai, K. and Fujii, K.},
  title = {Variational quantum algorithm for nonequilibrium steady states},
  journal = {Phys. Rev. Res.},
  volume = {2},
  pages = {043289},
  year = {2020}
}

@article{watad2024,
  author = {Watad, T. M. and Lindner, N. H.},
  title = {Variational quantum algorithms for simulation of {L}indblad dynamics},
  journal = {Quantum Sci. Technol.},
  volume = {9},
  pages = {025015},
  year = {2024}
}

@article{chen2024,
  author = {Chen, H. and Gomes, N. and Niu, S. and de Jong, W. A.},
  title = {Adaptive variational simulation for open quantum systems},
  journal = {Quantum},
  volume = {8},
  pages = {1252},
  year = {2024}

}

@article{kamakari2022,
  author = {Kamakari, H. and Sun, S.-N. and Motta, M. and Minnich, A. J.},
  title = {Digital quantum simulation of open quantum systems using quantum imaginary-time evolution},
  journal = {PRX Quantum},
  volume = {3},
  pages = {010320},
  year = {2022}
}

@article{santos2025,
  author = {Santos, S. and Song, X. and Savona, V.},
  title = {Low-rank variational quantum algorithm for the dynamics of open quantum systems},
  journal = {Quantum},
  volume = {9},
  pages = {1620},
  year = {2025}
}

@article{gravina2024,
  author = {Gravina, L. and Savona, V.},
  title = {Adaptive variational low-rank dynamics for open quantum systems},
  journal = {Phys. Rev. Res.},
  volume = {6},
  pages = {023072},
  year = {2024}
}

@article{lebris2013,
  author = {Le Bris, C. and Rouchon, P.},
  title = {Low-rank numerical approximations for high-dimensional {L}indblad equations},
  journal = {Phys. Rev. A},
  volume = {87},
  pages = {022125},
  year = {2013}
}

@article{zeynali2026,
  author = {Zeynali, A. and Bakhshi, Z.},
  title = {Adaptive low-rank variational quantum algorithm for simulating dissipative dynamics in photosynthetic complexes},
  journal = {Sci. Rep.},
  volume = {16},
  pages = {31846},
  year = {2026}
}

@article{zoufal2023,
  author = {Zoufal, C. and Sutter, D. and Woerner, S.},
  title = {Error bounds for variational quantum time evolution},
  journal = {Phys. Rev. Appl.},
  volume = {20},
  pages = {044059},
  year = {2023}
}

@article{medvidovic2023,
  author = {Medvidovi{\'c}, M. and Sels, D.},
  title = {Variational quantum dynamics of two-dimensional rotor models},
  journal = {PRX Quantum},
  volume = {4},
  pages = {040302},
  year = {2023}
}

@article{barison2021,
  author = {Barison, S. and Vicentini, F. and Carleo, G.},
  title = {An efficient quantum algorithm for the time evolution of parameterized circuits},
  journal = {Quantum},
  volume = {5},
  pages = {512},
  year = {2021}
}

@article{mitarai2019,
  author = {Mitarai, K. and Fujii, K.},
  title = {Methodology for replacing indirect measurements with direct measurements},
  journal = {Phys. Rev. Res.},
  volume = {1},
  pages = {013006},
  year = {2019}
}

@article{wang2009,
  author = {Wang, X. and Schirmer, S. G.},
  title = {Contractivity of the {H}ilbert-{S}chmidt distance under open-system dynamics},
  journal = {Phys. Rev. A},
  volume = {79},
  pages = {052326},
  year = {2009}
}

@article{perezgarcia2006,
  author = {P{\'e}rez-Garc{\'i}a, D. and Wolf, M. M. and Petz, D. and Ruskai, M. B.},
  title = {Contractivity of positive and trace-preserving maps under {$L_p$} norms},
  journal = {J. Math. Phys.},
  volume = {47},
  pages = {083506},
  year = {2006}
}

@book{nielsen2010,
  author = {Nielsen, M. A. and Chuang, I. L.},
  title = {Quantum Computation and Quantum Information},
  publisher = {Cambridge University Press},
  address = {Cambridge},
  year = {2010},
  edition = {10th Anniversary}
}

@article{cai2023,
  author = {Cai, Z. and Babbush, R. and Benjamin, S. C. and Endo, S. and Huggins, W. J. and Li, Y. and McClean, J. R. and O'Brien, T. E.},
  title = {Quantum error mitigation},
  journal = {Rev. Mod. Phys.},
  volume = {95},
  pages = {045005},
  year = {2023}
}

@article{qutip1,
  author = {Johansson, J. R. and Nation, P. D. and Nori, F.},
  title = {{QuTiP}: An open-source {P}ython framework for the dynamics of open quantum systems},
  journal = {Comput. Phys. Commun.},
  volume = {183},
  pages = {1760--1772},
  year = {2012}
}

@article{qutip2,
  author = {Johansson, J. R. and Nation, P. D. and Nori, F.},
  title = {{QuTiP} 2: A {P}ython framework for the dynamics of open quantum systems},
  journal = {Comput. Phys. Commun.},
  volume = {184},
  pages = {1234--1240},
  year = {2013}
}

@misc{qiskit2024,
  author = {Javadi-Abhari, A. and Treinish, M. and Krsulich, K. and Wood, C. J. and Lishman, J. and Gacon, J. and Martiel, S. and Nation, P. D. and Bishop, L. S. and Cross, A. W. and Johnson, B. R. and Gambetta, J. M.},
  title = {Quantum computing with {Q}iskit},
  howpublished = {arXiv:2405.08810},
  year = {2024}
}

@article{zhang2026,
  author = {Zhang, F. and Gomes, N. and Aftergood, J. and Iadecola, T. and Yao, Y. and Orth, P. P.},
  title = {Sampling noise and optimized measurement distribution in imaginary-time quantum dynamics simulations},
  journal = {Advanced Quantum Technologies},
  volume = {9},
  number = {9},
  pages = {e70418},
  archivePrefix = {arXiv},
  year = {2026}
}

@misc{arrasmith2020,
  author = {Arrasmith, A. and Cincio, L. and Somma, R. D. and Coles, P. J.},
  title = {Operator sampling for shot-frugal optimization in variational algorithms},
  howpublished = {arXiv:2004.06252},
  year = {2020}
}

@article{wecker2015,
  author = {Wecker, D. and Hastings, M. B. and Troyer, M.},
  title = {Progress towards practical quantum variational algorithms},
  journal = {Phys. Rev. A},
  volume = {92},
  pages = {042303},
  year = {2015}
}

@article{cerezo2021,
  author = {Cerezo, M. and Arrasmith, A. and Babbush, R. and Benjamin, S. C. and Endo, S. and Fujii, K. and McClean, J. R. and Mitarai, K. and Yuan, X. and Cincio, L. and Coles, P. J.},
  title = {Variational quantum algorithms},
  journal = {Nat. Rev. Phys.},
  volume = {3},
  pages = {625--644},
  year = {2021}
}

@article{bharti2022,
  author = {Bharti, K. and Cervera-Lierta, A. and Kyaw, T. H. and Haug, T. and Alperin-Lea, S. and Anand, A. and Degroote, M. and Heimonen, H. and Kottmann, J. S. and Menke, T. and Mok, W.-K. and Sim, S. and Kwek, L.-C. and Aspuru-Guzik, A.},
  title = {Noisy intermediate-scale quantum algorithms},
  journal = {Rev. Mod. Phys.},
  volume = {94},
  pages = {015004},
  year = {2022}
}

@article{li2017,
  author = {Li, Y. and Benjamin, S. C.},
  title = {Efficient variational quantum simulator incorporating active error minimization},
  journal = {Phys. Rev. X},
  volume = {7},
  pages = {021050},
  year = {2017}
}

@article{mcardle2019,
  author = {McArdle, S. and Jones, T. and Endo, S. and Li, Y. and Benjamin, S. C. and Yuan, X.},
  title = {Variational ansatz-based quantum simulation of imaginary time evolution},
  journal = {npj Quantum Inf.},
  volume = {5},
  pages = {75},
  year = {2019}
}

@article{dalibard1992,
  author = {Dalibard, J. and Castin, Y. and M{\o}lmer, K.},
  title = {Wave-function approach to dissipative processes in quantum optics},
  journal = {Phys. Rev. Lett.},
  volume = {68},
  pages = {580--583},
  year = {1992}
}

@article{plenio1998,
  author = {Plenio, M. B. and Knight, P. L.},
  title = {The quantum-jump approach to dissipative dynamics in quantum optics},
  journal = {Rev. Mod. Phys.},
  volume = {70},
  pages = {101--144},
  year = {1998}
}

@article{daley2014,
  author = {Daley, A. J.},
  title = {Quantum trajectories and open many-body quantum systems},
  journal = {Adv. Phys.},
  volume = {63},
  pages = {77--149},
  year = {2014}
}

@article{verstraete2004,
  author = {Verstraete, F. and Garc{\'i}a-Ripoll, J. J. and Cirac, J. I.},
  title = {Matrix product density operators: Simulation of finite-temperature and dissipative systems},
  journal = {Phys. Rev. Lett.},
  volume = {93},
  pages = {207204},
  year = {2004}
}

@article{zwolak2004,
  author = {Zwolak, M. and Vidal, G.},
  title = {Mixed-state dynamics in one-dimensional quantum lattice systems: A time-dependent superoperator renormalization algorithm},
  journal = {Phys. Rev. Lett.},
  volume = {93},
  pages = {207205},
  year = {2004}
}

@article{finazzi2015,
  author = {Finazzi, S. and Le Boit{\'e}, A. and Storme, F. and Baksic, A. and Ciuti, C.},
  title = {Corner-space renormalization method for driven-dissipative two-dimensional correlated systems},
  journal = {Phys. Rev. Lett.},
  volume = {115},
  pages = {080604},
  year = {2015}
}

@article{nagy2019,
  author = {Nagy, A. and Savona, V.},
  title = {Variational quantum {M}onte {C}arlo method with a neural-network ansatz for open quantum systems},
  journal = {Phys. Rev. Lett.},
  volume = {122},
  pages = {250501},
  year = {2019}
}

@article{vicentini2019,
  author = {Vicentini, F. and Biella, A. and Regnault, N. and Ciuti, C.},
  title = {Variational neural-network ansatz for steady states in open quantum systems},
  journal = {Phys. Rev. Lett.},
  volume = {122},
  pages = {250503},
  year = {2019}
}

@article{hartmann2019,
  author = {Hartmann, M. J. and Carleo, G.},
  title = {Neural-network approach to dissipative quantum many-body dynamics},
  journal = {Phys. Rev. Lett.},
  volume = {122},
  pages = {250502},
  year = {2019}
}

@article{yoshioka2019,
  author = {Yoshioka, N. and Hamazaki, R.},
  title = {Constructing neural stationary states for open quantum many-body systems},
  journal = {Phys. Rev. B},
  volume = {99},
  pages = {214306},
  year = {2019}
}

@article{kliesch2011,
  author = {Kliesch, M. and Barthel, T. and Gogolin, C. and Kastoryano, M. and Eisert, J.},
  title = {Dissipative quantum {C}hurch-{T}uring theorem},
  journal = {Phys. Rev. Lett.},
  volume = {107},
  pages = {120501},
  year = {2011}
}

@article{childs2017,
  author = {Childs, A. M. and Li, T.},
  title = {Efficient simulation of sparse {M}arkovian quantum dynamics},
  journal = {Quantum Inf. Comput.},
  volume = {17},
  pages = {901--947},
  year = {2017}
}

@inproceedings{cleve2017,
  author = {Cleve, R. and Wang, C.},
  title = {Efficient quantum algorithms for simulating {L}indblad evolution},
  booktitle = {44th International Colloquium on Automata, Languages, and Programming (ICALP 2017)},
  series = {LIPIcs},
  volume = {80},
  pages = {17:1--17:14},
  year = {2017}
}

@article{sweke2015,
  author = {Sweke, R. and Sinayskiy, I. and Bernard, D. and Petruccione, F.},
  title = {Universal simulation of {M}arkovian open quantum systems},
  journal = {Phys. Rev. A},
  volume = {91},
  pages = {062308},
  year = {2015}
}

@article{hu2020,
  author = {Hu, Z. and Xia, R. and Kais, S.},
  title = {A quantum algorithm for evolving open quantum dynamics on quantum computing devices},
  journal = {Sci. Rep.},
  volume = {10},
  pages = {3301},
  year = {2020}
}

@article{schlimgen2021,
  author = {Schlimgen, A. W. and Head-Marsden, K. and Sager, L. M. and Narang, P. and Mazziotti, D. A.},
  title = {Quantum simulation of open quantum systems using a unitary decomposition of operators},
  journal = {Phys. Rev. Lett.},
  volume = {127},
  pages = {270503},
  year = {2021}
}

@article{delgado2025,
  author = {Delgado-Granados, L. H. and Krogmeier, T. J. and Sager-Smith, L. M. and Avdic, I. and Hu, Z. and Sajjan, M. and Abbasi, M. and Smart, S. E. and Narang, P. and Kais, S. and Schlimgen, A. W. and Head-Marsden, K. and Mazziotti, D. A.},
  title = {Quantum algorithms and applications for open quantum systems},
  journal = {Chem. Rev.},
  volume = {125},
  pages = {1823--1839},
  year = {2025}
}

@article{vanstraaten2021,
  author = {van Straaten, B. and Koczor, B.},
  title = {Measurement cost of metric-aware variational quantum algorithms},
  journal = {PRX Quantum},
  volume = {2},
  pages = {030324},
  year = {2021}
}

@article{kubler2020,
  author = {K{\"u}bler, J. M. and Arrasmith, A. and Cincio, L. and Coles, P. J.},
  title = {An adaptive optimizer for measurement-frugal variational algorithms},
  journal = {Quantum},
  volume = {4},
  pages = {263},
  year = {2020}
}

@misc{gu2021,
  author = {Gu, A. and Lowe, A. and Dub, P. A. and Coles, P. J. and Arrasmith, A.},
  title = {Adaptive shot allocation for fast convergence in variational quantum algorithms},
  howpublished = {arXiv:2108.10434},
  year = {2021}
}

@article{gacon2024,
  author = {Gacon, J. and Nys, J. and Rossi, R. and Woerner, S. and Carleo, G.},
  title = {Variational quantum time evolution without the quantum geometric tensor},
  journal = {Phys. Rev. Res.},
  volume = {6},
  pages = {013143},
  year = {2024}
}

@article{corcoles2021,
  author = {C{\'o}rcoles, A. D. and Takita, M. and Inoue, K. and Lekuch, S. and Minev, Z. K. and Chow, J. M. and Gambetta, J. M.},
  title = {Exploiting dynamic quantum circuits in a quantum algorithm with superconducting qubits},
  journal = {Phys. Rev. Lett.},
  volume = {127},
  pages = {100501},
  year = {2021}
}

@article{ates2012,
  author = {Ates, C. and Olmos, B. and Garrahan, J. P. and Lesanovsky, I.},
  title = {Dynamical phases and intermittency of the dissipative quantum {I}sing model},
  journal = {Phys. Rev. A},
  volume = {85},
  pages = {043620},
  year = {2012}
}

@article{ruskai1994,
  author = {Ruskai, M. B.},
  title = {Beyond strong subadditivity? {I}mproved bounds on the contraction of generalized relative entropy},
  journal = {Rev. Math. Phys.},
  volume = {6},
  pages = {1147--1161},
  year = {1994}
}

@article{lebris2015,
  author = {Le Bris, Claude and Rouchon, Pierre and Roussel, Julien},
  title = {Adaptive low-rank approximation and denoised {Monte Carlo} approach for high-dimensional {Lindblad} equations},
  journal = {Phys. Rev. A},
  volume = {92},
  pages = {062126},
  year = {2015}
}

@article{etienney2026,
  author = {Etienney, Paul-Louis and Robin, R{\'e}mi and Rouchon, Pierre},
  title = {A posteriori error estimates for the {Lindblad} master equation},
  journal = {Quantum},
  volume = {10},
  pages = {2031},
  year = {2026}
}

@article{jassem2026,
  author = {Jassem, Abdullah Issa},
  title = {A posteriori error analysis and rank-adaptive structure-preserving numerical integration of large-scale {Lindblad} matrix differential equations},
  journal = {Journal of Iraqi Al-Khwarizmi},
  volume = {10},
  number = {2},
  pages = {380--398},
  year = {2026},
}

\end{document}